\documentclass[conference]{IEEEtran}

\usepackage{graphicx}
\usepackage[caption=false,font=footnotesize]{subfig}
\usepackage{amsmath}
\usepackage{amssymb}
\usepackage{cite}
\usepackage[hyphens]{url}
\usepackage{booktabs}
\usepackage{array}
\usepackage{float}
\usepackage{siunitx}
\usepackage[font=footnotesize,labelfont=bf,justification=centering]{caption}

\usepackage[
letterpaper,
top=0.72in,
bottom=1.0in,
left=0.67in,
right=0.67in
]{geometry}

\begin{document}
% ============================================================

\title{Transformer is All You Need: Attention-Based 
Anomaly Detection and Classification in Inverter-Rich Power Systems}

\author{%
  \IEEEauthorblockN{Emad Abukhousa, Saman Zonouz,
                    and A.\,P.\,Sakis Meliopoulos}
  \IEEEauthorblockA{\textit{School of Electrical and Computer Engineering,
    Georgia Institute of Technology, Atlanta, GA, USA}\\
    \{emadak, szonouz6, sakis.m\}@gatech.edu}
}

\maketitle

% ============================================================
\begin{abstract}
% ============================================================
Inverter-based resources and IEC 61850 process-bus measurements introduce new protection challenges, including nontraditional fault behavior and measurement-domain cyber-physical attacks. This paper evaluates DL-Xformer, an attention-based Transformer classifier for multi-class fault and cyberattack diagnosis, side-by-side with Dynamic State Estimation-Based Protection (DSE-EBP) on identical high-fidelity electromagnetic-transient (EMT) streaming measurements from an IBR-rich power grid. The evaluation uses an 18-class taxonomy covering normal operation, 11 physical faults, and six measurement-domain attacks, including CT/PT ratio manipulation and GPS spoofing, sampled at 4.8\,kHz from synchronized upstream and downstream merging units. DSE-EBP detects all streaming anomalies in 0.417--1.660\,ms, with a mean detection time of 0.756\,ms, while DL-Xformer classifies the same events in 2.50--50.42\,ms, with a mean classification time of 13.46\,ms. The longest delay occurs in a deliberate stress case where a CT ratio attack is introduced while residual oscillations from a preceding DLG fault have not fully settled; the event-window accuracy drops to 76.1\%, but the stable final classification remains correct. Measurement-level feature attribution shows that the DL-Xformer decision is driven by physically meaningful current and voltage channels at the attacked measurement location. Together, the two methods motivate a layered protection architecture for next-generation inverter-dominated smart grids.
\end{abstract}


% ============================================================
\section{Introduction}
% ============================================================

Modern power grids are transitioning toward high penetrations of
inverter-based resources (IBRs), including photovoltaic generation,
battery energy storage, and wind turbines.
Unlike synchronous machines, IBRs limit and reshape fault-current
contributions through fast control loops, producing lower magnitudes
and non-traditional transients that challenge conventional protection
assumptions~\cite{Nimpitiwan2007,Reno2021part1,Telukunta2017}.
Traditional overcurrent and distance protection schemes rely on
fault currents substantially exceeding rated values; in IBR-rich
systems, fault contributions can be as low as 1.1--1.5\,p.u.,
making threshold-based detection unreliable.
Bidirectional power flows from distributed generation further
disrupt directional coordination schemes.
Industry disturbance analyses document protection misoperations
in IBR-rich systems~\cite{NERC2017BlueCut}, confirming that this
is a structural, not transitional, challenge for the protection
engineering community.

At the same time, digital substations implementing IEC~61850
Process Bus architectures expose the measurement layer to
cyber-physical attacks.
CT ratio manipulation, PT ratio manipulation, and timing-reference
spoofing can corrupt sampled-value streams while preserving
plausible IEC~61850 frame timing and communication
behavior~\cite{Akbarzadeh2023,Moussa2018,Musleh2021}.
Unlike packet-level intrusions, these attacks may not violate
any protocol rule while corrupting the physical meaning of the
sampled values.
Moreover, a skilled adversary can manipulate CT ratios, PT ratios,
or timing references so that the resulting waveform resembles a
legitimate physical fault, masks a true fault, or creates spatial
inconsistency between upstream and downstream measurements.
The protection problem is therefore not only anomaly detection;
it is fault-versus-attack discrimination and localization of the
corrupted measurement source.
Detecting these attacks requires evaluating both waveform semantics
and physics consistency, which motivates model-based approaches
alongside data-driven classifiers.

Machine-learning and deep-learning methods have demonstrated
strong performance on fault and cyberattack classification from
power system waveforms~\cite{Hink2014,Khaw2021,Roy2023,
Oelhaf2025review}.
However, prior ML-based protection studies often report high
offline accuracy without evaluating continuous streaming behavior
under event-to-event interaction, confidence-gated abstention,
or delayed class stabilization.
This distinction is operationally important: a classifier can
eventually issue the correct event label while still exhibiting
reduced event-window accuracy if the correct class stabilizes
late in the event window.
In contrast, Dynamic State Estimation-Based Protection~(DSE-EBP)
detects measurement inconsistency through a physics-based
residual and therefore responds rapidly, but its output is an
anomaly or protection decision rather than a semantic
cyber-physical diagnosis.
This paper focuses on this latency-specificity gap: fast
physics-based detection versus slower but richer data-driven
diagnosis.

The authors' prior ensemble-learning study introduced a
high-fidelity EMT dataset for power-system fault and cyberattack
classification and showed that measurement-domain attacks can be
distinguished from physical faults in IBR-rich
waveforms~\cite{Abukhousa2025wisdom,Abukhousa2025dataset}.
The latency-aware benchmark then compared multiple deep-learning
architectures under streaming inference, revealing that
classification latency and event-window coverage are rarely
quantified for protection-grade
deployment~\cite{Abukhousa2025latency}.
The present work narrows the focus to the Transformer family and evaluates whether attention-based temporal modeling can provide reliable multi-class diagnosis, confidence behavior, and measurement-level attribution when compared side-by-side with DSE-EBP on identical high-fidelity EMT streams.
In parallel, DSE-EBP provides fast physics-based anomaly
detection by testing measurement consistency against a
protected-zone model~\cite{Meliopoulos2013,Meliopoulos2017,
Abukhousa2025PESGM}.
DSE-EBP is fast and model-grounded but does not provide the same
semantic label set as a trained diagnostic classifier.

Existing studies on IBR-rich protection often examine physics-based
relays and data-driven classifiers separately, with many emphasizing
offline accuracy rather than streaming diagnostic behavior. This leaves
an important gap in understanding how attention-based diagnosis and
physics-based protection compare when evaluated on the same
high-fidelity EMT measurements. This paper addresses that gap by
evaluating DL-Xformer and DSE-EBP side-by-side using detection time,
classification time, event-window accuracy, confidence evolution, and
measurement-level attribution.

\textit{Contributions:}
(1)~Evaluation of DL-Xformer for 18-class fault and measurement-domain attack diagnosis from 4.8\,kHz sampled values.
(2)~Side-by-side comparison with DSE-EBP on identical EMT streams, separating detection time from classification time.
(3)~Analysis of the latency--diagnosis tradeoff between fast physics-based detection and attention-based classification.
(4)~Measurement-level attribution linking the DL-Xformer decision to physically meaningful attacked-location channels.

% ============================================================
\begin{figure}[!t]
\centering
\includegraphics[width=0.95\columnwidth]{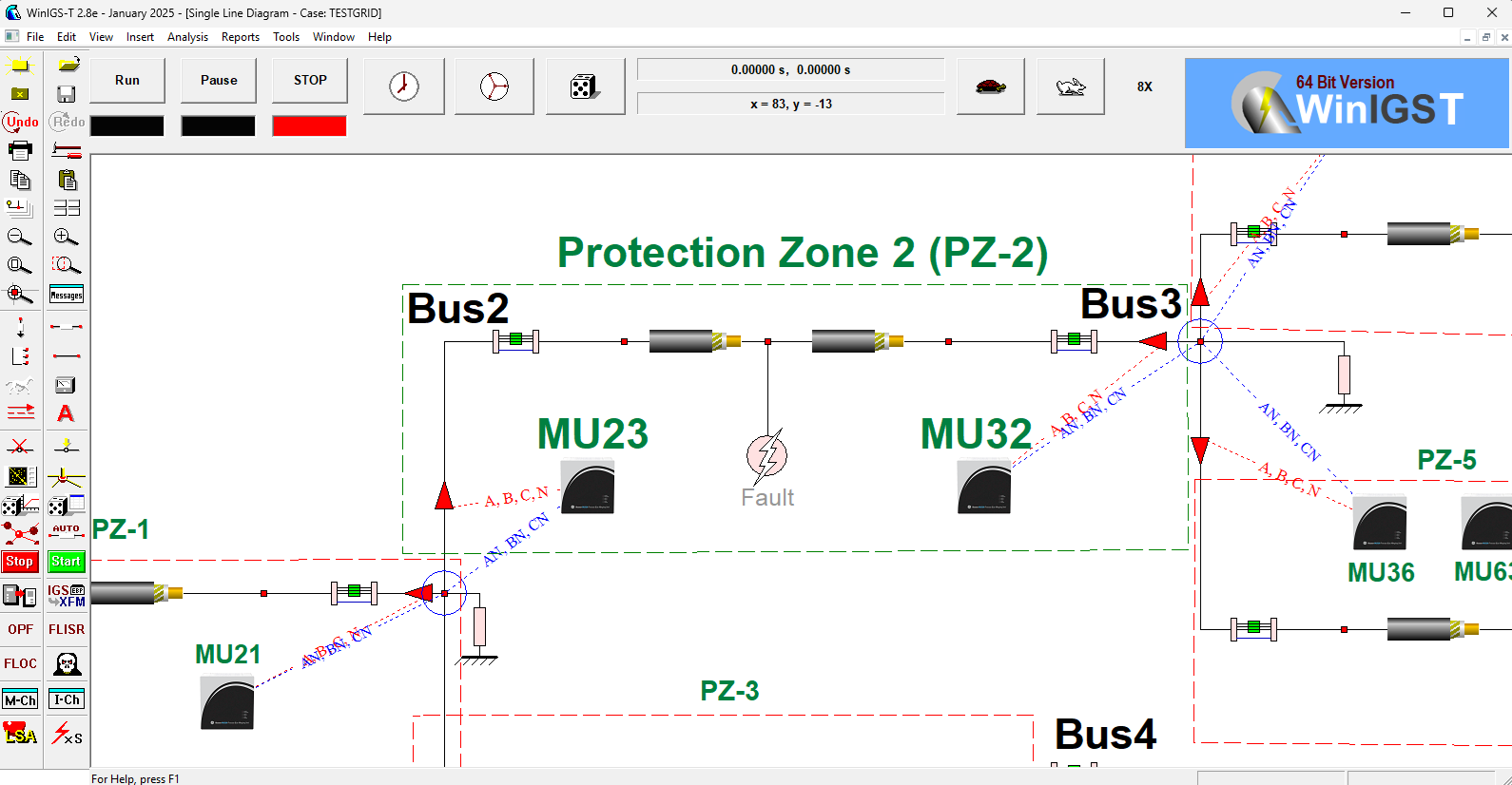}
\caption{EMT-modeled inverter-rich microgrid testbed generated in WinIGS.}
\label{fig:testbed}
\end{figure}

\section{Methodology}
% ============================================================

\subsection{High-Fidelity Testbed}

The high-fidelity electromagnetic-transient data were generated
using WinIGS~\cite{WinIGS}, an EMT simulation environment
developed by Advanced Grounding Concepts and cited in the
reference list, following the testbed and evaluation framework
of the authors' prior
studies~\cite{Abukhousa2025wisdom,Abukhousa2025latency}.
The network includes a 50\,kVA photovoltaic inverter, a
30\,kVA battery energy storage system, distribution lines,
step-down transformers, and five protection zones.
This paper studies the zone bounded by upstream MU23 and
downstream MU32 (Fig.~\ref{fig:testbed}).
Each MU digitizes three phase currents, neutral current, and
three phase-to-neutral voltages at $f_s = 4{,}800$\,Hz,
yielding $N_\mathrm{cyc} = 80$ samples per 60\,Hz cycle and
the 14-dimensional feature vector
\begin{equation}
  \mathbf{x}(i) =
    [\mathbf{I}_{23}(i),\,\mathbf{V}_{23}(i),\,
     \mathbf{I}_{32}(i),\,\mathbf{V}_{32}(i)]^\top
    \in \mathbb{R}^{14},
  \label{eq:feature}
\end{equation}
where $\mathbf{I}_{23},\mathbf{I}_{32}\in\mathbb{R}^{4}$ contain
the three phase currents and neutral current, and
$\mathbf{V}_{23},\mathbf{V}_{32}\in\mathbb{R}^{3}$ contain the
three phase-to-neutral voltages from MU23 and MU32, respectively.

\subsection{Training Dataset Generation}

The training data are drawn from the authors' previously
introduced high-fidelity EMT dataset for power-system fault
and cyberattack classification~\cite{Abukhousa2025wisdom,
Abukhousa2025dataset}, generated by simulating 22\,s of EMT
scenarios covering all 18 anomaly classes (Table~\ref{tab:taxonomy}).
Pre-processing: (i)~remove missing-value samples;
(ii)~exclude timestamp to prevent leakage;
(iii)~standardize with training-set statistics only;
(iv)~encode class labels; and
(v)~extract overlapping windows ($L=50$, stride~1), yielding
shape $(N_\mathrm{seq},50,14)$.
The 50-sample window is shorter than one 60\,Hz cycle and is
selected to preserve sub-cycle responsiveness while providing
sufficient waveform context for phase and neutral-current
spatial relationships.
The dataset is split 70\%~training / 10\%~validation /
20\%~test using stratified block-based partitioning;
class weights are applied to mitigate imbalance.

\begin{table}[!t]
\centering
\scriptsize
\caption{Training Dataset: 18-Class Anomaly Taxonomy}
\label{tab:taxonomy}
\begin{tabular}{@{}p{1.6cm}p{0.9cm}p{4.6cm}@{}}
\toprule
\textbf{Group} & \textbf{Classes} & \textbf{Description} \\
\midrule
Normal       & 0         & Normal operation \\
SLG faults   & 1--3      & A-N, B-N, C-N \\
LL faults    & 5--7      & A-B, A-C, B-C \\
DLG faults   & 9--11     & AB-N, AC-N, BC-N \\
3-ph faults  & 14--15    & ABC and ABC-N \\
CT attacks   & 4, 8      & CT ratio manip.\ at MU32, MU23 \\
PT attacks   & 12, 13    & PT ratio manip.\ at MU32, MU23 \\
GPS attacks  & 16, 17    & GPS spoofing at MU32, MU23 \\
\bottomrule
\multicolumn{3}{@{}l}{\scriptsize
  1 normal + 11 physical fault + 6 cyberattack = 18 classes.}
\end{tabular}
\end{table}

\subsection{Streaming Evaluation Scenario}

The streaming evaluation uses a 6-second prediction sequence
with five 200-ms events separated by normal operating intervals:
SLG A--N at 1.0\,s, LL B--C at 2.0\,s, DLG AC--N at 3.0\,s,
CT ratio attack at MU32 at 4.0\,s, and PT ratio attack at MU23
at 5.0\,s.
This sequence is generated from waveforms not used in training,
ensuring no data leakage between training and streaming evaluation.
The DLG AC--N event is intentionally placed immediately before
the CT attack so that the CT classification case is evaluated
under residual post-fault oscillation.
This is a deliberate simulation stress case, not a normal field
protection timeline; in field operation a protection trip would
typically isolate the fault before a later cyberattack develops
on top of the post-fault transient.

\subsection{Algorithmic Pipeline}

The complete evaluation pipeline has two branches operating on
the same synchronized measurement stream.
The \emph{data-driven branch} normalizes each 14-channel window
using the training-set scaler, applies DL-Xformer to produce
class probabilities, smooths over one cycle, and emits either
a class label or a confidence-gated abstention.
The \emph{physics-based branch} applies DSE-EBP to the same
MU23/MU32 measurements, computes the chi-square residual
statistic, converts it to a confidence level, and applies the
area-based trip logic.
The two branches do not feed into one another; they are evaluated
side-by-side on identical measurement streams, so any observed
performance difference is attributable solely to the method.

\subsection{DL-Xformer Diagnostic Classifier}

DL-Xformer is an attention-based neural network for diagnostic
event classification, not to be confused with physical power
transformers.
It operates on sliding 50-sample waveform windows and follows
the scaled dot-product attention mechanism
of~\cite{Vaswani2017}.
Given input $X\in\mathbb{R}^{L\times d}$, where $L=50$ is the
window length and $d$ is the model feature dimension, the model
computes query, key, and value matrices:
\begin{align}
  Q &= XW_Q, \label{eq:Q} \\
  K &= XW_K, \label{eq:K} \\
  V &= XW_V, \label{eq:V}
\end{align}
where $W_Q$, $W_K$, and $W_V$ are learned projection matrices.
Scaled dot-product attention is applied~\cite{Vaswani2017}:
\begin{equation}
  \mathrm{Attn}(Q,K,V)
    = \mathrm{softmax}\!\left(\frac{QK^\top}{\sqrt{d_k}}\right)V,
  \label{eq:attn}
\end{equation}
where $d_k$ is the key/query dimension for each head.
Multi-head attention concatenates $H$ parallel heads:
\begin{equation}
  \mathrm{MHA}(X) = \mathrm{Concat}(h_1,\ldots,h_H)W_O,
  \label{eq:mha}
\end{equation}
where $h_1,\ldots,h_H$ are the attention-head outputs and $W_O$
is the learned output projection.
Sinusoidal positional encoding preserves sample ordering:
\begin{align}
  PE(p,2j)   &= \sin\!\left(\tfrac{p}{10000^{2j/d}}\right),
  \label{eq:pe_sin} \\
  PE(p,2j+1) &= \cos\!\left(\tfrac{p}{10000^{2j/d}}\right),
  \label{eq:pe_cos}
\end{align}
where $p$ is the sample position inside the window and $j$
indexes the embedding dimension.
Four stacked attention blocks followed by global average pooling,
a dense/dropout layer, and softmax produce a class probability
vector $P(i)\in[0,1]^{18}$.
A one-cycle centered moving-average stabilizes streaming
decisions:
\begin{equation}
  \bar{P}_k(i)
    = \frac{1}{N_\mathrm{cyc}}
      \sum_{j=i-N_\mathrm{half}}^{i+N_\mathrm{half}} P_k(j),
  \quad N_\mathrm{cyc}=80,\;N_\mathrm{half}=40,
  \label{eq:smooth}
\end{equation}
where $P_k(j)$ is the softmax probability assigned to class $k$
at streaming index $j$.
The confidence-gated decision is
\begin{equation}
  y(i) =
  \begin{cases}
    \arg\max_k \bar{P}_k(i), & \max_k \bar{P}_k(i) \ge \tau, \\
    -1,                      & \max_k \bar{P}_k(i) < \tau,
  \end{cases}
  \label{eq:gate}
\end{equation}
with $\tau=0.60$.
The centered filter introduces a look-ahead delay of
$T_\mathrm{smooth} = N_\mathrm{half}/f_s \approx 8.33$\,ms;
a causal trailing-window implementation is required for
relay-grade deployment and is left for future work.

\subsection{DSE-EBP Detection Method}

DSE-EBP represents the protected zone through a physics-based
model and detects anomalies by testing whether measured currents
and voltages are consistent with the protected-zone physical
model~\cite{Meliopoulos2013,Meliopoulos2017,Abukhousa2025PESGM}.
The weighted least-squares~(WLS) residual objective is
\begin{equation}
  J(x) = [h(x) - z]^\top W [h(x) - z],
  \label{eq:wls}
\end{equation}
where $z$ is the measurement vector, $h(x)$ is the nonlinear
model-predicted measurement vector, $x$ is the state vector,
and $W$ is the measurement-weight matrix.
The observed chi-square residual statistic is
\begin{equation}
  \xi(t) = \sum_m \!\left(
    \frac{h_m(\hat{x}(t)) - z_m(t)}{k\sigma_m}
  \right)^{\!2},
  \label{eq:chisq}
\end{equation}
where $\hat{x}(t)$ is the estimated state, $z_m(t)$ is
measurement $m$, $h_m(\hat{x}(t))$ is the corresponding
model-predicted measurement, $\sigma_m$ is the measurement-error
standard deviation, and $k$ is the desensitization factor.
The confidence level is
\begin{equation}
  CL(t) = \Pr\bigl\{\chi^2_\nu \ge \xi(t)\bigr\},
  \label{eq:cl}
\end{equation}
where $\chi^2_\nu$ is a chi-square random variable with
$\nu = m - n$ degrees of freedom ($m$ measurements, $n$ states).
Note that $\xi(t)$ is the observed residual statistic plotted as
the \(\chi^2\) signal in the figures.
A larger $\xi(t)$ indicates poorer model-measurement consistency,
causing $CL(t)$ to decrease; anomaly detection appears as a
rise in $\xi(t)$ and a simultaneous collapse of $CL(t)$.

The relay accumulates low-confidence area over reset window $T_r$:
\begin{equation}
  a(t) = \int_{t-T_r}^{t} \max\bigl\{0,\,1-CL(s)\bigr\}\,ds,
  \label{eq:area}
\end{equation}
and issues a trip when the accumulated area exceeds threshold
$T_d$:
\begin{equation}
  \mathrm{Trip}(t) =
  \begin{cases}
    1, & a(t) > T_d, \\
    0, & a(t) \le T_d.
  \end{cases}
  \label{eq:trip}
\end{equation}
Relay parameters: $T_r = 100$\,ms, $T_d = 40$\,ms, $k = 1.0$.
DSE-EBP detection time is the interval from event onset to
the first chi-square elevation and confidence collapse.
The trip signal follows approximately 40\,ms after onset,
governed by (\ref{eq:area})--(\ref{eq:trip}).
\subsection{Evaluation Metrics}

\textit{DSE-EBP detection time} is measured from event onset to the first sustained rise in the plotted chi-square statistic \(\chi^2(t)\) with confidence collapse below 80\%. \textit{DL-Xformer classification time} is measured from event onset to the first sustained correct class label after confidence gating; it is not a trip time. \textit{Event-window accuracy} is the fraction of samples in the 200\,ms event window assigned the correct label after confidence gating, so lower accuracy can reflect delayed stabilization rather than wrong final diagnosis. \textit{Stable final classification} is the final confident class label after one-cycle probability smoothing and thresholding by \(\tau\); otherwise the model abstains.
% ============================================================
\section{Results}
% ============================================================

\begin{figure}[!t]
\centering
\includegraphics[width=0.98\columnwidth]{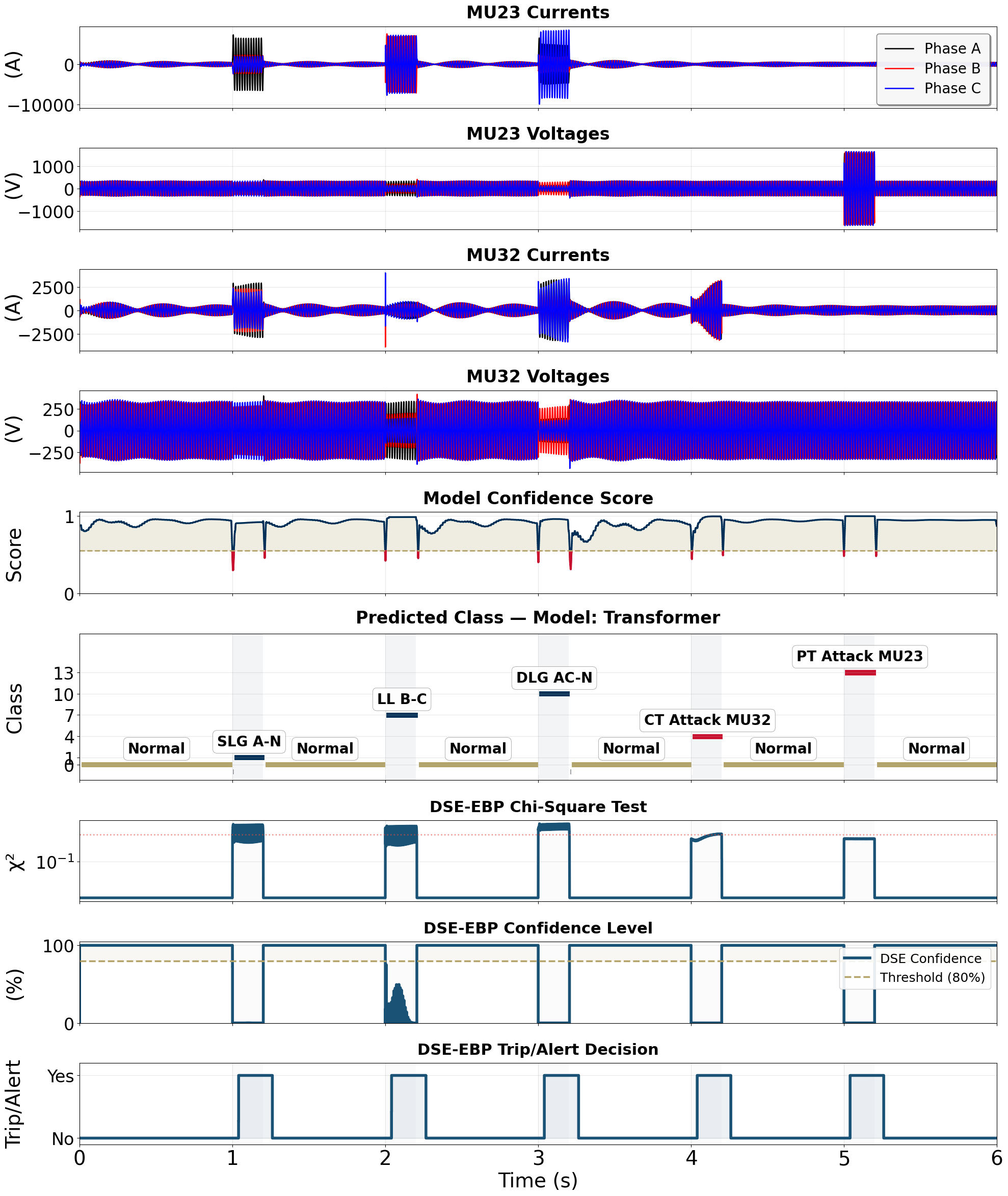}
\caption{Full five-event streaming evaluation side-by-side.
  Top: raw MU23/MU32 measurements.
  Middle: DL-Xformer confidence and predicted class.
  Bottom: DSE-EBP observed chi-square statistic~$\xi(t)$
  (plotted as $\chi^2$), confidence level~$CL(t)$, and trip
  decision. Both systems detect all five anomalies.}
\label{fig:eventstream}
\end{figure}
\subsection{Full Streaming Evaluation}

Fig.~\ref{fig:eventstream} shows the full five-event streaming
evaluation on the same MU23/MU32 measurement stream.
The upper panels contain synchronized current and voltage
waveforms; the middle panels show DL-Xformer confidence and
predicted class; the lower panels show the DSE-EBP observed
chi-square statistic~$\xi(t)$ (plotted as the $\chi^2$ signal),
confidence level~$CL(t)$, and trip decision.
Both methods operate on the same physical disturbance sequence
but produce different outputs: DSE-EBP produces a fast
physics-consistency detection signal, while DL-Xformer produces
a semantic event label and confidence trajectory.

On held-out non-streaming test windows, DL-Xformer achieved
97.47\% accuracy across the 18-class taxonomy.
In the streaming sequence, performance is evaluated through
confidence behavior, classification time, and event-window
accuracy.
DL-Xformer provides stable diagnostic labels for the SLG, LL,
DLG, and PT events.
The CT attack case is intentionally harder because it starts
after the DLG AC--N event, before the waveform fully settles.
DSE-EBP responds to each anomaly through a sharp rise in
$\xi(t)$ and a collapse of $CL(t)$, followed by the area-based
trip signal approximately 40\,ms after onset, consistent with
$T_d = 40$\,ms.
This establishes the main comparison: DSE-EBP detects quickly;
DL-Xformer provides event-specific diagnosis.

\subsection{Detection Time, Classification Time, and
            Event-Window Accuracy}

\begin{figure}[!t]
\centering
\subfloat[Classification time per event.\label{fig:time}]{%
  \includegraphics[width=0.96\columnwidth]{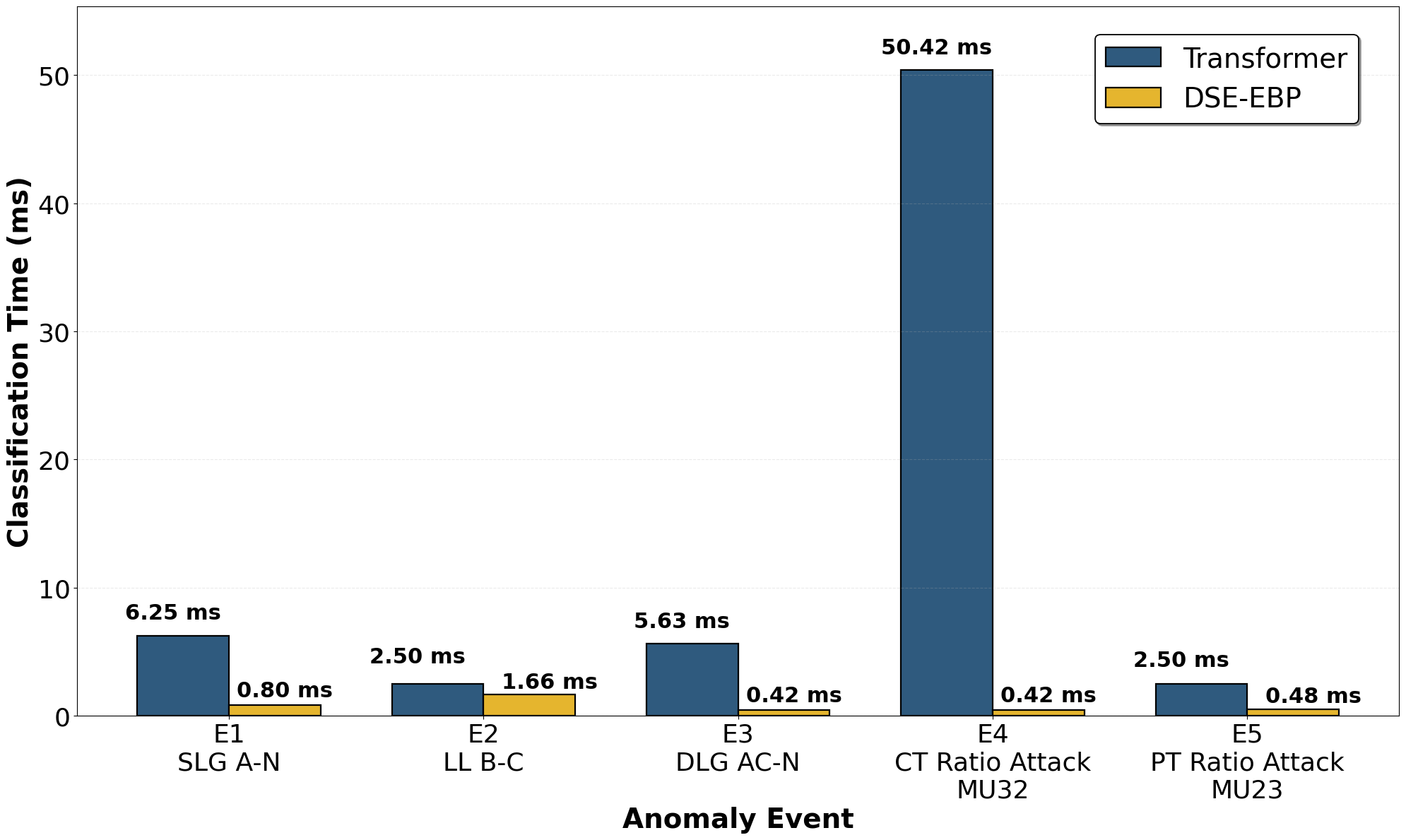}}\\[2mm]
\subfloat[Event-window accuracy per event.\label{fig:acc}]{%
  \includegraphics[width=0.96\columnwidth]{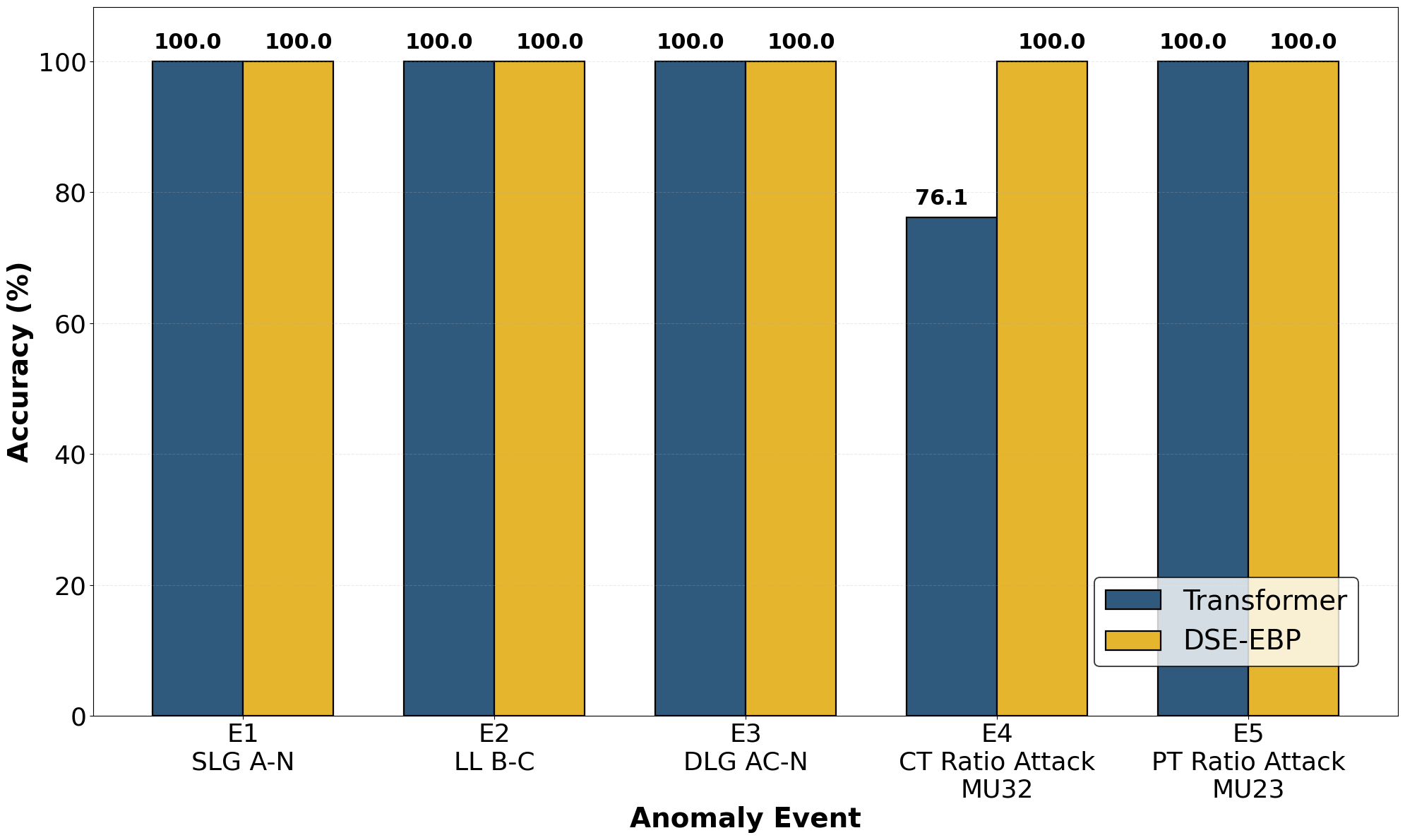}}
\caption{Timing and event-window accuracy comparison.}
\label{fig:timeacc}
\end{figure}

DSE-EBP detects all five streaming anomalies in
0.417--1.660\,ms (mean 0.756\,ms), demonstrating suitability
for primary protection where inverter over-current shutdown
occurs within 2--5\,ms.
DL-Xformer classifies the same events in 2.50--50.42\,ms
(mean 13.46\,ms); the mean DL-Xformer classification time is
17.8 times larger than the mean DSE-EBP detection time,
reflecting the difference between fast anomaly detection and
semantic event classification.
For Events~1, 2, 3, and~5 the DL-Xformer classification time
is within one power cycle (16.67\,ms at 60\,Hz).
Event~4 (CT attack, 50.42\,ms) is the exception and is
analyzed in detail in Sections~\ref{sec:dlgcase}
and~\ref{sec:ctcase}.
The comparison is not a replacement claim: DSE-EBP is faster
for primary anomaly detection; DL-Xformer provides semantic
classification, attack localization, confidence behavior, and
measurement-level attribution.
Fig.~\ref{fig:timeacc} visualizes both metrics across all
five events.

\subsection{Zoomed Case A: DLG AC--N Fault}
\label{sec:dlgcase}

\begin{figure}[!t]
\centering
\includegraphics[width=0.98\columnwidth]{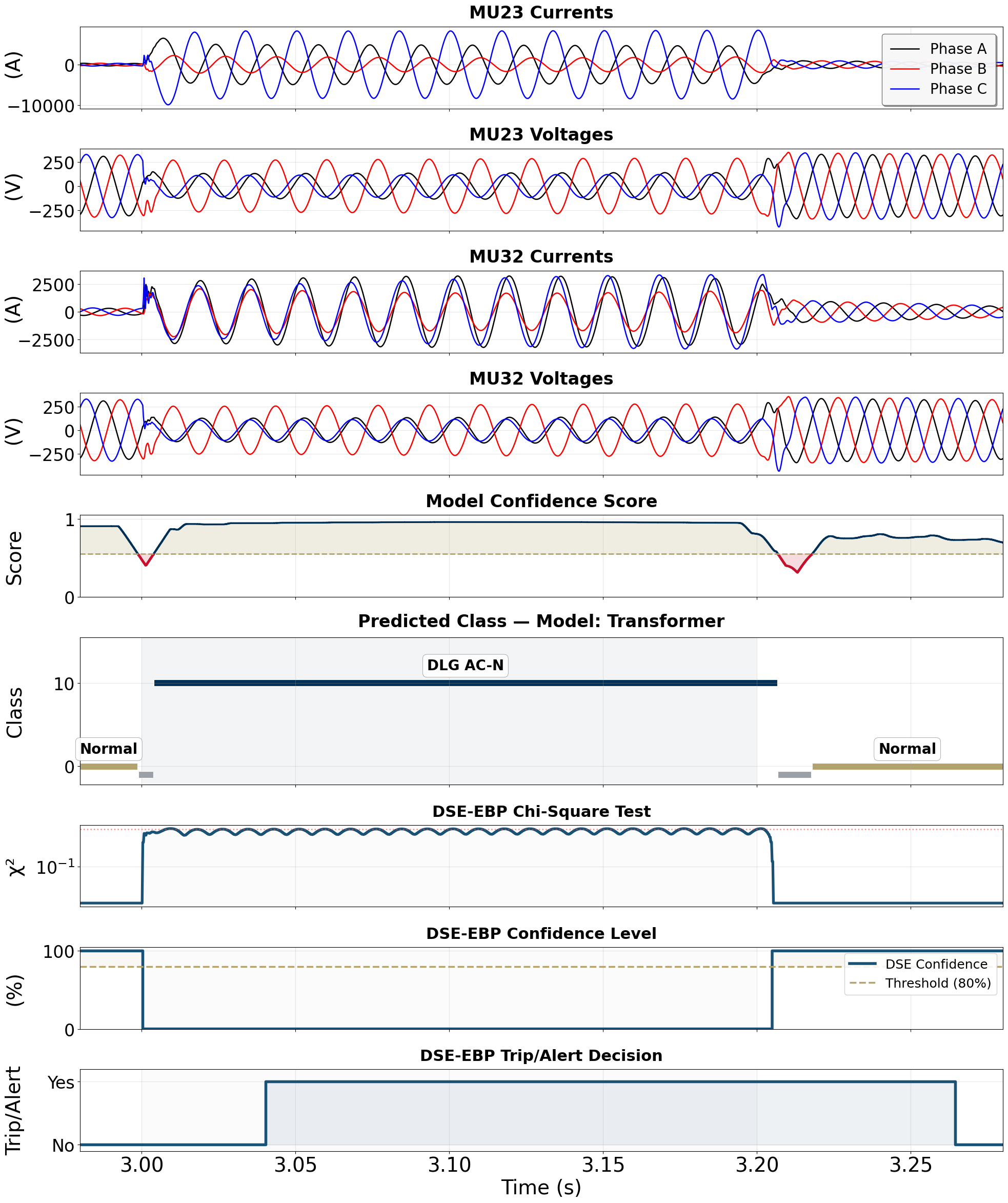}
\caption{Case~A: DLG AC--N fault at 3.0\,s.}
\label{fig:dlg}
\end{figure}

Fig.~\ref{fig:dlg} shows Event~3, the DLG AC--N fault at 3.0\,s.
Phases~A and~C and the neutral are involved, producing severe
oscillatory behavior and elevated neutral-current unbalance.
Both systems respond correctly: DSE-EBP detects in 0.418\,ms;
DL-Xformer classifies in 5.63\,ms.
The key significance is what this event leaves behind: oscillatory
transients and residual unbalance that do not fully settle before
Event~4 begins 800\,ms later, creating the most challenging
classification condition in the evaluation.

\subsection{Zoomed Case B: CT Attack Under Residual Oscillation}
\label{sec:ctcase}

\begin{figure}[!t]
\centering
\includegraphics[width=0.98\columnwidth]{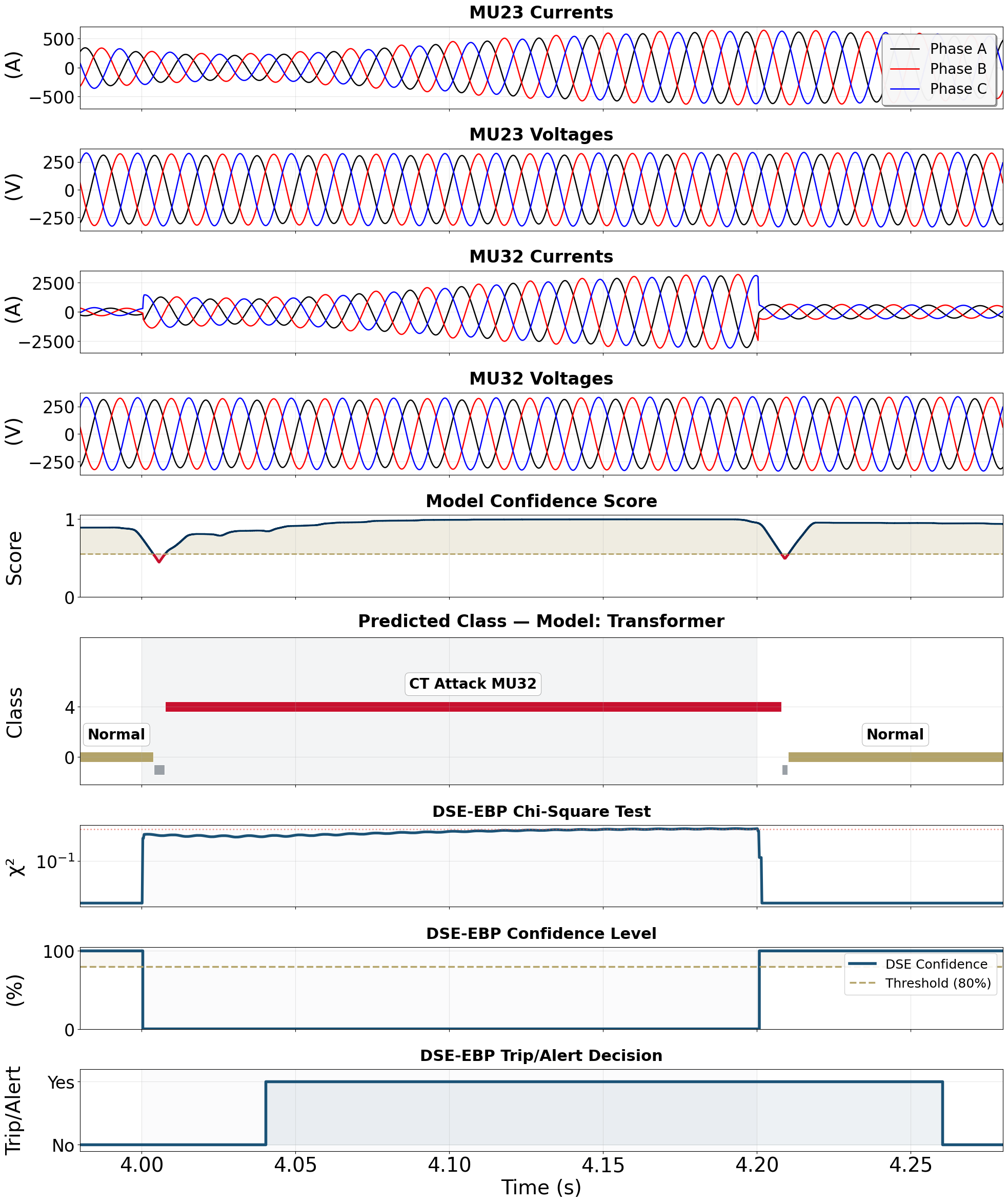}
\caption{Case~B: CT ratio attack on MU32 at 4.0\,s under residual
  DLG oscillation.}
\label{fig:ctcase}
\end{figure}

Fig.~\ref{fig:ctcase} shows the CT ratio attack on MU32 at 4.0\,s.
Upstream MU23 currents remain normal; MU32 current channels show
characteristic CT magnitude distortion.
The classification challenge arises because early DL-Xformer
windows carry a mixture of residual DLG signatures and the new
CT attack, requiring additional temporal evidence before the
CT attack signature becomes dominant and confidence exceeds~$\tau$.

The classification time is 50.42\,ms and event-window accuracy
is 76.1\%.
Both figures reflect a \emph{timing-window effect}, not an
incorrect final diagnosis: once DL-Xformer issues a stable
classification, the attack is correctly identified as the CT
ratio attack on MU32.
DSE-EBP detection time remains short at 0.417\,ms because CT
manipulation immediately violates the protected-zone current-balance
constraints, causing an instantaneous rise in $\xi(t)$ regardless
of prior system history.
This contrast directly illustrates the complementary strengths
of the two approaches.

\subsection{Measurement-Level Attribution}

\begin{figure}[!t]
\centering
\includegraphics[width=0.98\columnwidth]{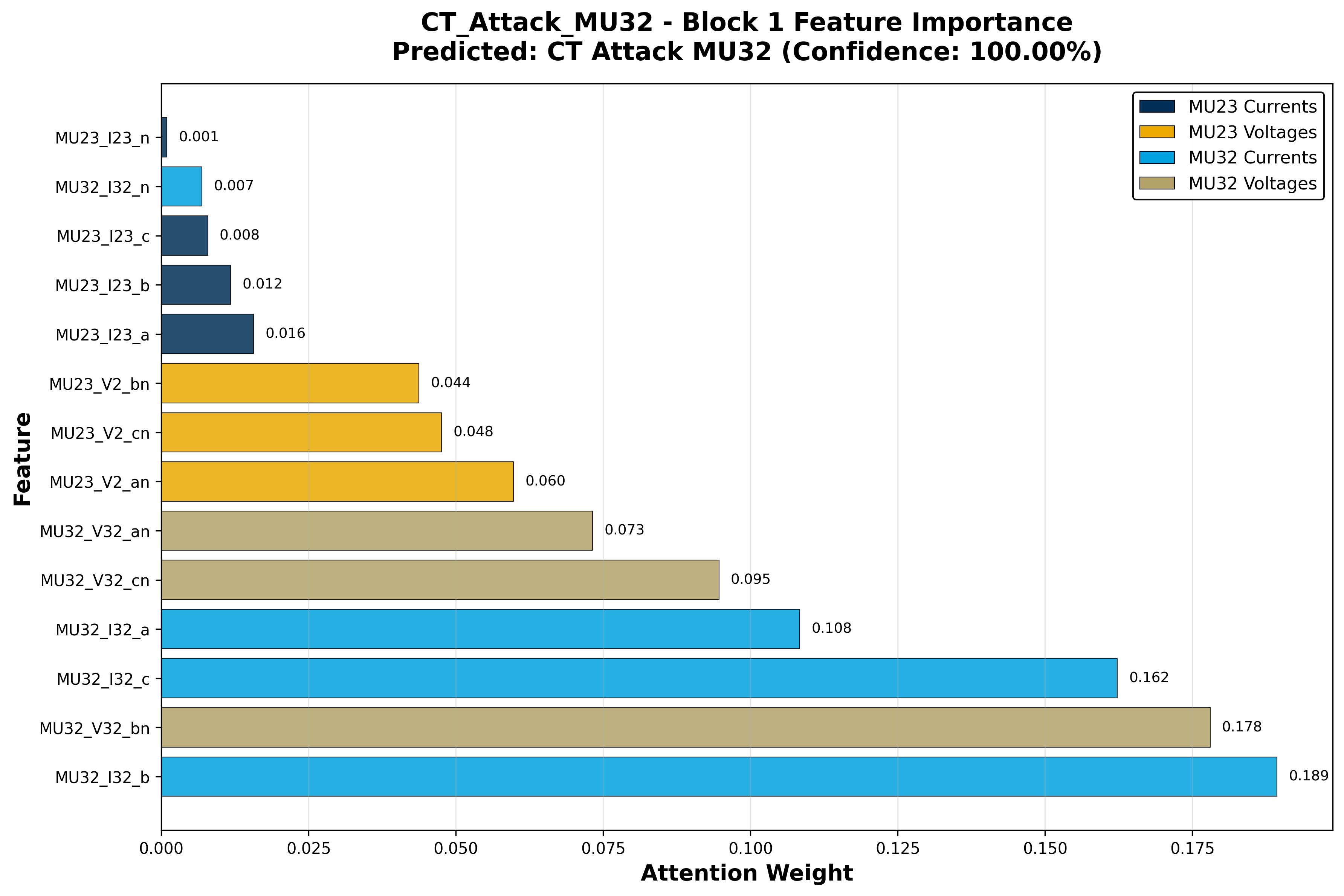}
\caption{Measurement-level feature-attribution scores for the
  CT ratio attack on MU32.
  Dominant weights concentrate on MU32 phase-current and
  voltage channels, the physically attacked signals.}
\label{fig:attrib}
\end{figure}

Fig.~\ref{fig:attrib} shows measurement-level feature-attribution
scores for the CT attack case.
Dominant attribution weights concentrate on MU32 phase-current
channels ($I_{32,a}$, $I_{32,b}$, $I_{32,c}$) and MU32 voltage
channels---the physically attacked measurement signals.
MU23 measurements, which are not manipulated in this scenario,
receive significantly lower attribution.
The MU32 neutral-current channel contributes as secondary context:
the preceding DLG AC--N fault creates residual neutral-current
unbalance that the model uses to distinguish the CT manipulation
from the earlier DLG waveform pattern.
This supports the interpretation that DL-Xformer learns
physically grounded spatial relationships between upstream and
downstream measurements without explicit physics supervision.

% ============================================================
\section{Discussion}
% ============================================================
\textit{Latency-specificity tradeoff.}
DSE-EBP provides fast physics-based anomaly detection and area-integrated trip support (0.756\,ms mean detection; $\sim$40\,ms trip), making it suitable for primary protection functions that must operate within a few cycles. DL-Xformer is slower (13.46\,ms mean classification) but provides diagnostic value DSE-EBP alone cannot supply: fault type, cyberattack type, localization, confidence behavior, and feature attribution. The two methods answer different operational questions: DSE-EBP supports fast protection action, while DL-Xformer supports diagnostic awareness.

\textit{Deployment interpretation.}
The side-by-side results motivate a layered protection framework in which DSE-EBP serves as the primary physics-based detection layer, while DL-Xformer serves as a secondary diagnostic and situational-awareness layer. In such a framework, DSE-EBP would provide the fast trigger when protected-zone measurement consistency is violated, while DL-Xformer would distinguish physical faults from measurement-domain attacks and localize corrupted sources. This work establishes complementary profiles needed for such a hybrid architecture as future work.

\textit{Field deployability and limitations.}
The current DL-Xformer implementation is not yet field-ready for primary protection. Its centered smoothing filter introduces 8.33\,ms of look-ahead, and worst-case CT attack classification requires 50.42\,ms under residual DLG oscillation. While useful for diagnostic support, this is not yet sufficient for protection requiring deterministic operation within two to three cycles. Future work should reduce deep-learning latency through causal filtering, shorter inference windows, and hardware acceleration. The evaluation is simulation-based and uses fixed attack magnitudes. Controller-hardware-in-the-loop and real-time digital simulation studies are needed before field deployment, along with broader feeder topologies, variable attack magnitudes, and relay coordination tests.

\section{Conclusion}

This paper presented a side-by-side evaluation of DL-Xformer, an attention-based diagnostic classifier, and physics-based Dynamic State Estimation-Based Protection (DSE-EBP) for cyber-physical security in inverter-rich power systems. Using identical high-fidelity EMT streaming measurements generated in the WinIGS testbed, DSE-EBP detected all five streaming anomalies within 0.417--1.660\,ms, with a mean detection time of 0.756\,ms, supporting its role as a fast primary anomaly-detection mechanism.  DL-Xformer classified the same events in 2.50--50.42\,ms, with a mean classification time of 13.46\,ms, while providing semantic event identification, attacked-measurement localization, confidence behavior, and measurement-level attribution. In the most challenging case, a CT ratio attack introduced under residual DLG oscillation, DL-Xformer reached the correct stable diagnosis despite a reduced 76.1\% event-window accuracy. The attribution results further showed reliance on physically meaningful current and voltage channels at the attacked measurement location. These results demonstrate the complementary strengths of fast physics-based anomaly detection and richer data-driven diagnosis. They motivate a future layered protection architecture in which DSE-EBP provides rapid primary detection and DL-Xformer provides diagnostic classification, localization, and interpretability for IBR-dominated grids.

% ============================================================
\bibliographystyle{IEEEtran}
\bibliography{references_master}
% ============================================================

\end{document}